\documentclass[twocolumn,prb,superscriptaddress,amsmath,amssymb,aps]{revtex4}
\usepackage{graphicx}
\usepackage{epsfig}
\usepackage{bm}
\usepackage{miller}
\usepackage{color}

\begin{document}
 
\title{Comparison of analytic and numerical bond-order potentials for W and Mo
\footnotetext{This is an author-created, un-copyedited version of an article accepted
for publication in J. Phys.: Condens. Matter. The publisher is not responsible
for any errors or omissions in this version of the manuscript or any version
derived from it. The Version of Record is available online at http://dx.doi.org/10.1088/0953-8984/25/26/265002.}
}

\author{M. \v{C}\'{a}k}
\author{T. Hammerschmidt}
\author{R. Drautz}
\affiliation{ Atomistic Modelling and Simulation, ICAMS, Ruhr-Universit\"{a}t Bochum, D-44801 Bochum, Germany  }

\date{\today}

\begin{abstract}
Bond-order potentials (BOPs) are derived from the tight-binding (TB) approximation and provide a linearly-scaling computation of the 
energy and forces for a system of interacting atoms. While the numerical BOPs involve the numerical integration of the response 
(Green's) function, the expressions for the energy and interatomic forces are analytical within the formalism of 
the analytic BOPs. In this paper we present a detailed comparison of numerical and analytic BOPs. We use 
established parametrisations for the bcc refractory metals W and Mo and test structural energy differences; 
tetragonal, trigonal, hexagonal and orthorhombic deformation paths; formation energies of point defects as well as phonon 
dispersion relations. We find that the numerical and analytic BOPs generally are in very good agreement for the calculation 
of energies. Different from the numerical BOPs, the forces in the analytic BOPs correspond exactly to the negative gradients
of the energy. This makes it possible to use the analytic BOPs in dynamical simulations and leads to improved predictions of defect
energies and phonons as compared to the numerical BOPs.
\end{abstract}

\maketitle

\section{Introduction}

Refractory metals exhibit properties that make them unique compared to other metals.
They have an exceptionally high melting point,
excellent strength at high temperatures as well as good wear, corrosion and abrasion resistance.
Moreover, they show very high hardness and good electrical and heat conducting properties. Here we focus on the bcc refractory metals W and Mo.
Tungsten and its alloys are used for light-bulbs and filaments and in the future possibly as plasma-facing wall material in fusion reactors. 
Molybdenum is often used as alloying element in high-strength steels (up to 8\%) and in superalloys (Ni- or Co-based) 
in order to increase the melting temperature. One of the more recent applications are bulk metallic glasses with 
exceptionally high glass-transition and crystallisation temperatures.~\cite{Mo-Si-glasses}

Modelling the mechanical behaviour of refractory metals requires to account for microstructural defects, such as dislocations and grain boundaries as well
as secondary phase precipitates. Empirical potentials, such as Finnis-Sinclair~\cite{Finnis-84} or the embedded-atom method (EAM)~\cite{EAM}, only partly capture the nature of bonding
mediated by {\em d} electrons in bcc transition metals~\cite{Groeger_1}. Tight-binding (TB) methods, as approximate electronic structure methods, 
are attractive because they can treat large systems and at the same time offer
an understanding of the underlying physical processes. The bond-order potentials (BOPs) provide a linear scaling approximate 
solution to the TB problem~\cite{Pettifor-89,Aoki-93-2,Horsfield-96,Drautz-06,Drautz_Pettifor_2011,Hammerschmidt-09-1,Hammerschmidt-09-2,Finnis-book}. 
BOPs have proved to be particularly effective for simulating 
properties of dislocations~\cite{Girshick-98-1,Girshick-98-2,Vitek2004_MSE,Mrovec-04,Mrovec-07-2,Katzarov2007,Gröger2008_I,Katzarov2009} or in 
explaining the origin of brittle cleavage in iridium~\cite{Cawkwell-05-2,Cawkwell-06}.

While the properties of dislocations predicted from BOPs are typically more robust and reliable than predictions from empirical potentials, 
a recent study~\cite{Cereceda-2013} deemed numerical BOPs as not suitable for finite temperature dynamical simulations because
of a mismatch between the forces and the negative gradients of the energy. Analytic BOPs~\cite{Drautz-06,Drautz_Pettifor_2011,Bernhard_KPM},
in contrast, provide forces that exactly match the negative gradients of the energy. 

In this paper we present a detailed comparison between analytic and numerical BOPs. We show that the energies predicted using  
analytic BOPs and  numerical BOPs are essentially equivalent, such that existing parametrisations of numerical BOPs may directly be used in
dynamical simulations with the analytic BOPs. In section~~\ref{sec:Methodology} we briefly review the main differences between the numerical and
analytic BOPs. In section~\ref{sec:Comparison} the BOPs are compared by using the electronic density of states, structural energy differences, point 
defect formation energies, structural transformation paths and phonon spectra.

\section{Methodology}
\label{sec:Methodology}
\subsection{Tight Binding}

The derivation of BOPs starts from the TB approximation that expresses the eigenfunctions $\psi_n$ of the Schr{\"o}dinger equation
\begin{equation}
\label{eq:Schroedinger}
\hat{H}\psi_n = E_n \psi_n \, ,
\end{equation}
in a minimal basis of orbitals $\alpha$ centred on atoms $i$ 
\begin{equation}
|\psi_n \rangle= \sum\limits_{i\alpha} c_{i\alpha}^{(n)} |i\alpha\rangle \, .
\end{equation}
For an orthonormal basis, the eigenvalues $E_n$ and the coefficients 
$c_{i\alpha}^{(n)}$ of the eigenfunctions $\psi_n$ are determined by solving the secular equation
\begin{equation}
\label{eq:secularEq}
\sum\limits_{j\beta} H_{i\alpha j\beta} c_{j\beta}^{(n)} = E_n c_{i\alpha}^{(n)} \, .
\end{equation}
The matrix elements $H_{i\alpha j\beta}$ are typically expressed as functions of the interatomic distance. 
The diagonalisation of the Hamiltonian matrix
\begin{equation}
\label{eq:HamMatrix}
H_{i\alpha j\beta} = \langle i\alpha | \hat{H} | j\beta \rangle \, ,
\end{equation}
is the computationally most demanding part of TB calculations. In the TB bond model~\cite{Sutton-88,Drautz_Pettifor_2011}, the binding energy of
a d-valent, charge neutral and non-magnetic material is given as the sum over the covalent bond energy $U_{\mathrm{bond}}$ 
and the repulsive energy $U_{\mathrm{rep}}$ 
\begin{equation}
\label{eq:TBBM}
U_B = U_{\mathrm{bond}} + U_{\mathrm{rep}} \, . 
\end{equation}
The bond energy $U_{\mathrm{bond}}$ can be expressed in onsite 
and intersite representation. Both representations are equivalent but offer different views on bond formation. 
The onsite representation is based on the 
atom-based local density of states $n_{i\alpha}$ on atom $i\alpha$,
\begin{equation}
n_{i\alpha} (E) = \sum\limits_{n} \left|c_{i\alpha}^{(n)}\right|^2 \delta (E-E_n) \, .
\end{equation}
The intersite representation is expressed in terms of the bond-order $\Theta_{i\alpha j\beta}$ or the density matrix 
$\rho_{i\alpha j\beta}$ between orbital $\alpha$ on atom $i$ and orbital $\beta$ on atom $j$ and is given by the sum over occupied states
\begin{equation}
\label{eq:bondorder}
\Theta_{i\alpha j\beta} = 2 \rho_{i\alpha j\beta} = 2 \sum\limits_{n}^{\mathrm{occ}} c_{i\alpha}^{*(n)} c_{j\beta}^{(n)} \, .
\end{equation}
The bond energy in onsite and intersite representation is given by 
\begin{eqnarray}
\label{eq:TB-Ubond}
U_{\mathrm{bond}} &=& 2 \sum\limits_{i\alpha} \int\limits^{E_{\mathrm{F}}} \left( E-E_{i\alpha} \right) n_{i\alpha}(E)dE  \\ \nonumber
                  &=& \sum\limits_{i\alpha \ne j\beta} \Theta_{i\alpha j\beta} H_{i\alpha j\beta} \, ,
\end{eqnarray}
where $E_{\mathrm{F}}$ is the Fermi level and $E_{i\alpha}$ = $H_{i\alpha i\alpha}$ are the diagonal elements of the Hamiltonian matrix. 
An in-depth discussion and interpretation of the bond order for molecules and solids is given in Refs.~\onlinecite{Finnis-book,Pettifor-95}.

\subsection{Bond-Order Potentials}

The {\em p}-th moment of the local density of states $n_{i\alpha}(E)$ is given by~\cite{Ducastelle_moments,Horsfield-96}
\begin{eqnarray}
\label{eq:momDOS}
\mu^{(p)}_{i\alpha} &=& \int E^{p} n_{i\alpha}(E)dE = \langle i\alpha | \hat{H}^{p} | i\alpha \rangle \\ \nonumber
                    &=& \sum\limits_{j_{1}\beta_{1}, j_{2}\beta_{2}...j_{p-1}\beta_{p-1}} H_{i\alpha j_{1}\beta_{1}} H_{j_{1}\beta_{1} j_{2}\beta_{2}}... H_{j_{p-1}\beta_{p-1}i\alpha} \, .
\end{eqnarray}
Using the last equality one understands the {\em p}-th moment of the local density of states as a closed loop of {\em p} hops along 
neighbouring atomic sites.
The local density of states $n_{i\alpha}(E)$ can be reconstructed from its moments $\mu^{(p)}_{i\alpha}$ by making use of the recursion method~\cite{Haydock1,Haydock2} 
with the on-site Green's function $G_{00}(E)$ expressed as a continued fraction
\begin{equation}
\label{eq:contfrac}
 G_{00}(E)=\cfrac{1}{E-a_0-\cfrac{b_1^2}{\ddots-\cfrac{\ddots}{E-a_\infty-\cfrac{b_\infty^2}{\ddots}}}}
\end{equation}
with recursion coefficients $a_{i}$ and $b_{i}$. The recursion coefficients may be computed from the moments of the density of states. Typically, 
for a single band, one calculates the first few recursion coefficients, equivalent to the first $m_{\rm max}$ moments, and estimates the following
recursion coefficients as $a_{\infty}$ and $b_{\infty}$, independent of $i$. Because the part of the continued fraction that involves only 
$a_{\infty}$ and $b_{\infty}$ can be evaluated to a square-root analytically~\cite{Horsfield-96}, this is referred to as the square-root terminator.

The local density of states is related to the Green's function~\cite{Horsfield-96}
\begin{equation}
\label{eq:GFdos}
n_{i\alpha}(E)=-\frac{1}{\pi}\lim_{\epsilon \to 0}\mathrm{Im}\{G_{i\alpha i\alpha}(E+i\epsilon)\} \, .
\end{equation} 
From Eq.~\ref{eq:TB-Ubond} one can then calculate the bond energy by numerical integration, which is referred to as numerical BOP. The forces
are obtained approximately by using the Hellmann-Feynman theorem. The numerical integration is one of the computational bottlenecks of such calculations.
In numerical BOP, an effective electronic temperature $T_{e}$ is introduced in order to improve the convergence of the bond-order expansion and 
to obtain a better agreement between the approximate Hellmann-Feynman forces and the true forces, i.e., the negative gradients of the energy~\cite{Horsfield-96,Girshick-98-1}.
As the introduction of the electronic temperature is an approximation on top of the numerical BOP expansion, in the following we use an electronic
temperature of $k_{B}T_{e} = $ 0.001~eV to keep the influence on binding energies and forces as small as possible. This value is smaller than the typically 
used $k_{B}T_{e}=$ 0.3~eV, but still provides numerical stability and very good agreement of the bond energies from numerical and analytic BOPs.

In the analytic BOPs, the density of states is expanded using Chebyshev polynomials of the second kind $U_{m}(\epsilon)$~\cite{Drautz-06,Drautz_Pettifor_2011},
\begin{eqnarray}
 n_{i\alpha}(\epsilon) & = & \frac{2}{\pi}\sqrt{1-\epsilon^{2}}\Bigg[\sum\limits_{m=0}^{m_{\rm{max}}}g_{U}^{(m)}\sigma_{i\alpha}^{(m)}U_{m}(\epsilon) \\ \nonumber
                       &   & + \sum\limits_{m_{\rm{max}}+1}^{m_{\rm{exp}}}g_{U}^{(m)}\sigma_{i\alpha}^{(m)}U_{m}(\epsilon)\Bigg] \, . 
\end{eqnarray}
The expansion coefficients $\sigma_{i\alpha}^{(m)}$ are obtained from the moments ${\mu}_{i\alpha}^{(p)}$. In analogy to the
numerical BOPs, only the first few expansion coefficients corresponding to $m_{\rm max}$ moments are explicitly computed. The remaining 
expansion coefficients up to $m_{\rm exp}$ are obtained from the square-root terminator~\cite{Bernhard_KPM}. The terminator coefficients 
$a_{\infty}$ and $b_{\infty}$ are also used to ensure that the density of states is contained in the band $-1\le \epsilon \le +1$, with 
$\epsilon = \frac{E-a_{\infty}}{2b_{\infty}}$. We approximate the values of $a_{\infty}$ and $b_{\infty}$ from the upper and lower bounds of
the energy spectra,
\begin{equation}
E_{\rm{min}} = a^{\rm{min}} - 2b^{\rm{max}}  \quad {\rm and} \quad  E_{\rm{max}} = a^{\rm{max}} - 2b^{\rm{max}} \, ,
\end{equation}
and therefore
\begin{equation}
 a_{\infty} = \frac{E_{\rm{max}}+E_{\rm{min}}}{2} \quad {\rm and} \quad b_{\infty} = \frac{E_{\rm{max}}-E_{\rm{min}}}{4} \, .
\end{equation}
The damping factors $g_{U}^{(m)}$ vary smoothly from one at $m = 0$ to zero at $m = m_{\rm exp}$ (see Fig.~2 in Ref.~\onlinecite{Bernhard_KPM}) and prevent Gibbs ringing in the expansion such that
the resulting density of states is always strictly positive~\cite{Bernhard_KPM,Kernel_Polynomials_Jackson}.

\subsection{Functional form and parametrisation}
\label{sec:FuncForm}

Our comparison between analytic and numerical BOPs is based on previously developed parametrisations for Mo~\cite{Mrovec-04} and W~\cite{Mrovec-07-2}.
The bond energy $U_{\mathrm{bond}}$ (Eq.~\ref{eq:TB-Ubond}) in the d-valent TB model is determined by the matrix elements $H_{i\alpha j\beta}$ (Eq.~\ref{eq:HamMatrix})
that are expressed in terms of two-centre Slater-Koster~\cite{Slater-54} bond integrals $\beta(R)$ and parametrised by the Goodwin-Skinner-Pettifor (GSP) function~\cite{GSP_function}
\begin{equation}
\label{eq:GSP}
\beta(R)=\beta(R_{0})\Big(\frac{R_{0}}{R}\Big)^{n_{a}}\mathrm{exp}\Big\{n_{b}\Big[\Big(\frac{R_{0}}{R_{c}}\Big)^{n_{c}}-\Big(\frac{R}{R_{c}}\Big)^{n_{c}}\Big]\Big\} \, .
\end{equation}
where $R_{0}$ is the first nearest neighbour distance in bcc. The long-range tail of the GSP function is smoothly forced to zero by a cut-off 
function between $R_{\rm{tail}}$ and $R_{\rm{cut}}$. In the interval $R_{\rm{cut}}$ - $R_{\rm{tail}}$ the GSP function is 
replaced by a fifth-order polynomial that guarantees continuous second derivatives. The number of d electrons $N_{d}$ is taken as 
$N_{d}$ = 4.2 for both, Mo and W.
An environmental repulsive term $U_{\mathrm{rep}}^{\mathrm{env}}$ is introduced in order to account for a correct description of 
the Cauchy pressure and is modelled by a Yukawa-like many-body environmentally dependent repulsive term~\cite{Duc-Cauchy-Pressure},  
\begin{equation}
\label{eq:Yukawa1}
U_{\mathrm{rep}}^{\mathrm{env}} = \frac{1}{2} \sum\limits_{i,i\neq j} \frac{B}{r_{ij}}\mathrm{exp}[-\lambda_{ij}(r_{ij}-2r_{\mathrm{core}})] \, ,
\end{equation}
where
\begin{equation}
\lambda_{ij}=\frac{1}{2}(\lambda_{i}+\lambda_{j}) \, ,  
\end{equation}
and
\begin{equation}
\label{eq:Yukawa3}
 \lambda_{i}=\lambda_{0} + \Big[\sum\limits_{k \neq i}C\mathrm{exp}(-\nu R_{ik})\Big]^{1/m} \, .
\end{equation}
Just like for the bonding integrals we are using a fifth-order polynomial as cutoff function for the Yukawa-like term
that acts on equations~\ref{eq:Yukawa1} and~\ref{eq:Yukawa3}. 
The pair potential term $U_{\mathrm{rep}}^{\mathrm{pair}}$ accounts for the repulsive short-range character of the atomic 
interactions. It is represented by a cubic spline,
\begin{equation}
\label{eq:Spline}
 U_{\mathrm{rep}}^{\mathrm{pair}} = \frac{1}{2}\sum\limits_{i,i \neq j} \sum\limits_{k=1}^{4}A_{k}(R_{k}-R_{ij})^{3} \, ,
\end{equation}
with parameters chosen such that the pair potential vanishes between the second and third nearest neighbour. For this reason,
no cut-off function needs to be applied.

\subsection{Computational details}

The calculations presented in the following were carried out with OXON~\cite{Horsfield-96,Horsfield-96-1} and BOPfox~\cite{BOPfox}. 
Both packages provide a TB kernel and we confirmed that the TB results using OXON and BOPfox are in excellent agreement. For the TB calculations 
presented in the following we used a Monkhorst-Pack k-point mesh~\cite{Monkhorst-Pack} and the tetrahedron method~\cite{tetrahedron} for integrating the 
Brillouin zone. For all TB calculations we used a k-point mesh of 30$\times$30$\times$30 which is sufficient also for the calculations
along the deformation paths, where the symmetry is lowered compared to bcc. 
For the BOP calculations, we used OXON for the numerical BOP and BOPfox for the analytic BOP calculations. Both BOPs 
use the same TB model with the functional form given in section~\ref{sec:FuncForm} and parameters given in Refs.~\onlinecite{Mrovec-04,Mrovec-07-2}. 
Therefore, the contributions of the repulsive energies are identical in analytic BOP (using BOPfox) and numerical BOP (using OXON), only 
the bond energy is treated with different formalisms.

\section{Comparison of analytic and numerical BOP}
\label{sec:Comparison}

\subsection{Density of states}

\begin{figure}[htb]
\centering
\begin{tabular}{c}
\epsfig{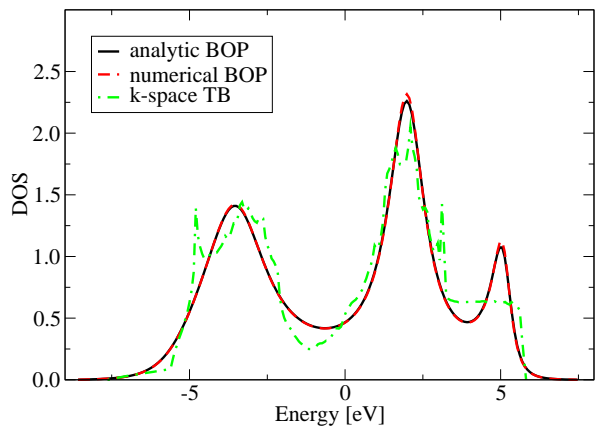}
\end{tabular}
\caption{Density of states of tungsten calculated using different approximations.}
\label{fig:DOS}
\end{figure}

In Fig.~\ref{fig:DOS}, we compare the density of states of bcc W as obtained with the numerical BOP using 9 moments and with the analytic BOP using $m_{\rm{max}}=9$ 
moments and $m_{\rm{exp}}=200$ to the TB reference calculations.
We find excellent agreement between the numerical BOP and the analytic BOP.
Nine moments in the BOP calculations are sufficient to reproduce the central features of the TB density of states, particularly
the positions of the bonding and anti-bonding peaks and the pseudo gap. This number of moments was also used in the original 
parametrisations~\cite{Mrovec-04,Mrovec-07-2} and previously shown to be sufficient for describing structural stability in 
transition metals~\cite{Bernhard_Thomas_TCP}.  The following tests were carried out with the same numbers of moments.

\subsection{Structural stability}
\label{sec:StructuralStability}

\begin{figure}[htb]
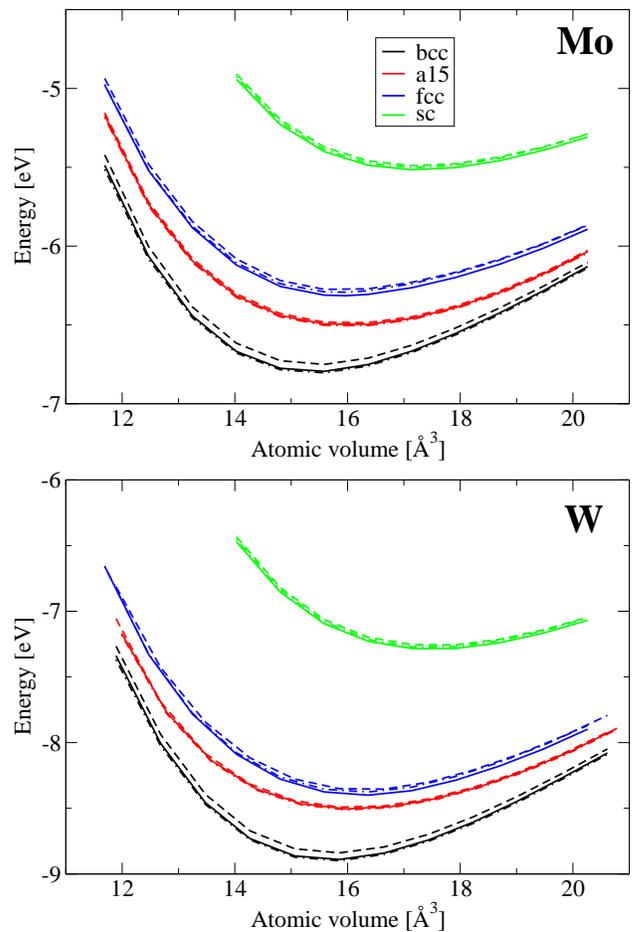

\centering
\begin{tabular}{c}
\epsfig{file=2.eps ,width=0.95\linewidth,clip=} \\
\epsfig{file=3.eps,width=0.95\linewidth,clip=} \\
\end{tabular}
\caption{Energy vs. volume curves. Full lines represent analytic BOP, dashed lines numerical BOP and dot-and-dash lines TB.}
\label{fig:EvsV}
\end{figure}

Experimental and calculated properties of the bcc ground state for Mo and W are summarised in Tab.~\ref{tab:BCC}.
\begin{table}[htb]
\begin{center}
\begin{tabular}{c c c c c c}
\hline\hline
   &        & expt   & analytic BOP & numerical BOP & TB \\
\hline
Mo   &  $C_{11}$ & 2.901 & 2.974 & 2.972 & 3.181 \\
     &  $C_{12}$ & 1.008 & 0.931 & 0.946 & 0.825 \\
     &  $C_{44}$ & 0.680 & 0.603 & 0.730 & 0.422 \\
     &  $E_{\rm{coh}}$ &  -6.82 & -6.79 & -6.75 & -6.80  \\
     &  $a_{0}$       & 3.147 &  3.147 & 3.147 &  3.147 \\
\hline 
 W   &  $C_{11}$  & 3.261  & 3.320 & 3.311 & 3.535 \\
     &  $C_{12}$  & 1.276  & 1.195 & 1.213 & 1.083 \\
     &  $C_{44}$  & 1.002  & 0.911 & 1.045 & 0.704\\
     &  $E_{\rm{coh}}$ & -8.90 & -8.89 & -8.84 &  -8.90  \\
     &  $a_{0}$    &  3.165 & 3.165 & 3.165 & 3.165 \\
   \hline\hline    
\end{tabular}
\caption{Elastic constants $C_{11}$, $C_{12}$ and $C_{44}$ [eV/\AA$^{3}$], cohesive energies [eV/atom] and lattice constants 
         [\AA] as obtained from experiment (see references in Ref.~\onlinecite{Finnis-84}), from analytic and numerical BOP and from TB.}
\label{tab:BCC}
\end{center}
\end{table}
The energies for analytic and numerical BOP as well as for TB as a function of atomic volume are shown in Fig.~\ref{fig:EvsV}. 
For both analytic and numerical BOP, the elastic constants are determined by fitting a fifth-order polynomial to the energy versus deformation data.
From Tab.~\ref{tab:BCC} one can see a good agreement between analytic and numerical BOP values of elastic parameters and cohesive energies with a 
slightly better match of the analytic BOP data to the TB reference than the numerical BOP.
Figure~\ref{fig:EvsV} shows that for both Mo and W the analytic and numerical BOPs are in a very good agreement, predicting
essentially the same energetics of the structures presented here, once more with a slightly better match of the TB data by
analytic BOP as compared to numerical BOP.

\subsection{Transformation paths}
\label{sec:TransformationPaths}

We consider several transformation or deformation paths in bcc. We calculate the energy as a function of the deformation parameter and compare 
it to TB. A more detailed description of the geometries of these paths can be found in literature~\cite{Paidar_deformations,orthorhombic_deformation}. 
Various deformation paths were studied in relation to the stability of the higher energy phases and extended 
defects~\cite{Wang-Sob-PRB-1999,Sob-Wang-Vitek-CMS-1997}.  

\subsubsection{Tetragonal deformation path}

\begin{figure}[htb]
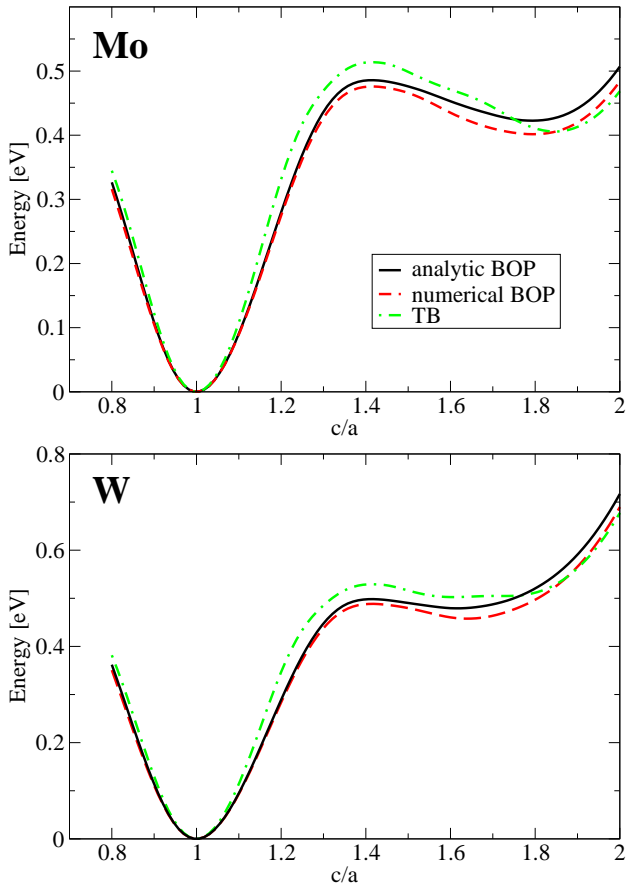

\centering
\begin{tabular}{c}
\epsfig{file=4.eps,width=0.95\linewidth,clip=} \\
\epsfig{file=5.eps,width=0.95\linewidth,clip=} \\
\end{tabular}
\caption{Binding energy along the tetragonal deformation path. The minimum at $c/a$ = 1 corresponds to bcc, the maximum 
         at $c/a$=$\sqrt2$ to fcc.}
\label{fig:DefTetra}
\end{figure}

The tetragonal deformation path follows loading of bcc along the [001] direction with the deformation parameter $c/a$. 
Here $c$ is the lattice parameter along [001] and $a$ along [100] a [010]. The volume
of the unit cell is conserved along this path. In a coordinate system with [001] and [100] parallel to the $z$ and 
$x$ axis, the only non-zero components of the Green-Lagrangian strain tensor for this deformation path are  
\begin{eqnarray}
  \label{eq:Tetra}
  \epsilon_{11}=\epsilon_{22}=\frac{a^{2}-a_{0}^{2}}{2a_{0}^{2}}, \quad \epsilon_{33}=\frac{c^{2}-a_{0}^{2}}{2a_{0}^{2}} \, ,
\end{eqnarray}
where $a_{0}$ is a lattice parameter of perfect bcc. Along this transformation path, $c/a$=1 and $c/a$=$\sqrt2$
correspond to bcc and fcc, respectively. These are visible as minimum (bcc) and maximum (fcc) in the binding energy
along the transformation path as compiled in Fig.~\ref{fig:DefTetra}.
The agreement between analytic and numerical BOP as well as the reference TB calculations is very good in the whole 
range of deformations. We note that the region around the global minimum (bcc) is related to the tetragonal shear modulus
$C^{'}$. Importantly, we find the correct positions and energies of the local maximum for fcc (symmetry dictated) and 
the local minimum (not dictated by symmetry) at $c/a$ = 1.6-1.8. 

\subsubsection{Trigonal deformation path}

\begin{figure}[htb]
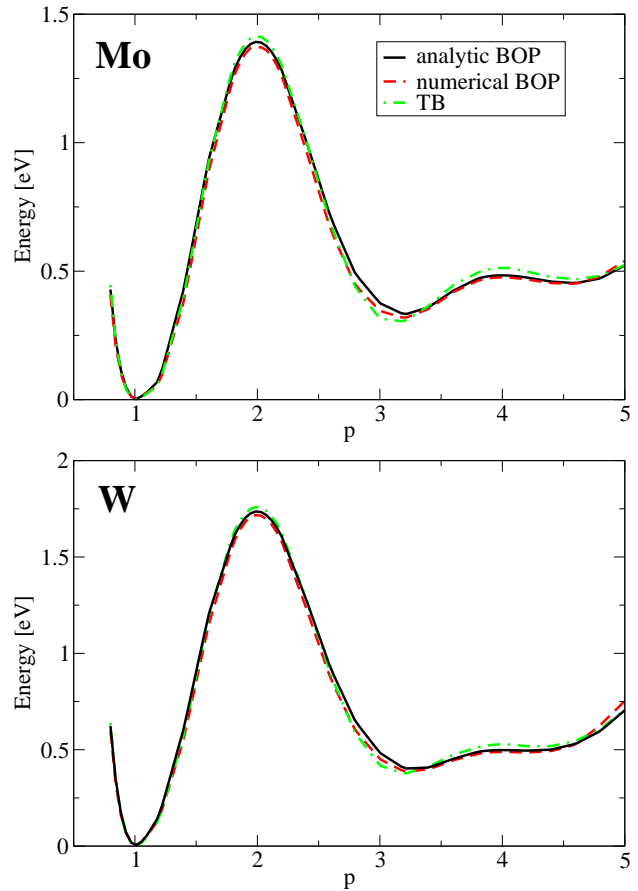

\centering
\begin{tabular}{c}
\epsfig{file=6.eps,width=0.95\linewidth,clip=} \\
\epsfig{file=7.eps ,width=0.95\linewidth,clip=} \\
\end{tabular}
\caption{Binding energy along the trigonal deformation path. The minimum at $p$ = $1$ corresponds to bcc, the maxima at $p$=$2$ and 
         $p$=$4$ to fcc and sc, respectively.}
\label{fig:DefTri}
\end{figure}

The trigonal deformation path represents a deformation of bcc with loading/compression along [111]. The atomic volume along the path
is conserved and the trigonal deformation connects bcc, sc and fcc at $p$=1, $p$=2 and $p$=4, respectively, see Fig.~\ref{fig:DefTri}. 
The agreement between analytic BOP, numerical BOP and TB is excellent along the deformation path including the local 
maximum at $p=4$. The curvature around the global energy minimum at $p$=1 is related to the trigonal (or rhombohedral) shear modulus $C_{44}$. 

\subsubsection{Hexagonal deformation path}

\begin{figure}[htb]
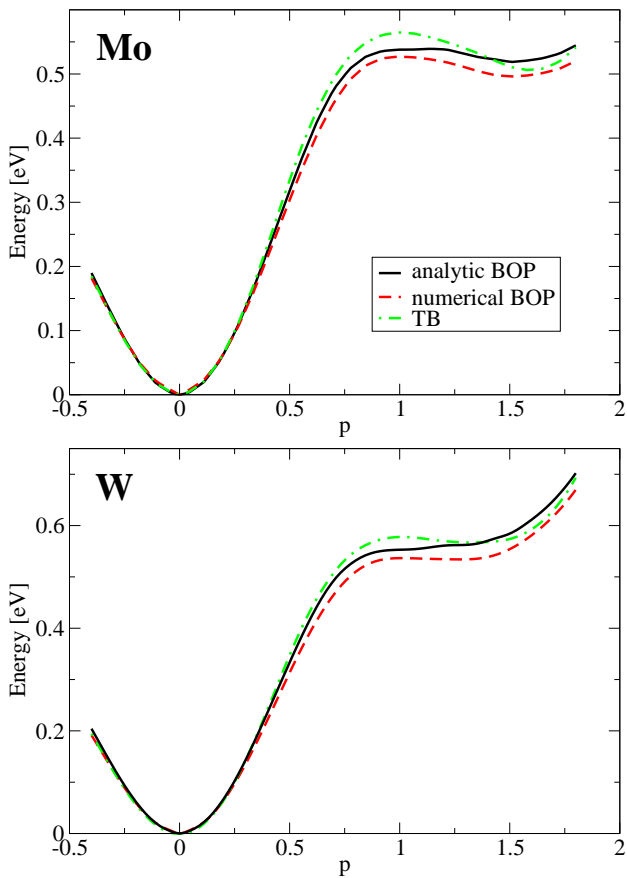

\centering
\begin{tabular}{c}
\epsfig{file=8.eps,width=0.95\linewidth,clip=} \\
\epsfig{file=9.eps ,width=0.95\linewidth,clip=} \\
\end{tabular}
\caption{Binding energy along the hexagonal deformation path. The minimum at $p$ = $0$ corresponds to bcc, the maximum at $p$=$1$ to hcp.}
\label{fig:DefHex}
\end{figure}

The hexagonal deformation path connects bcc with the hexagonal closed-packed (hcp) structure. It combines loading with a linearly 
coupled shuffling of the atomic planes~\cite{Paidar_deformations,Mrovec-04}.
In our representation, $p$ = 0 and $p$ = 1 represent bcc and hcp, respectively. From our results compiled in Fig.~\ref{fig:DefHex} 
we see that the agreement between analytic and numerical BOP and TB is very good along the full transformation path.

\subsubsection{Orthorhombic deformation path}

\begin{figure}[htb]
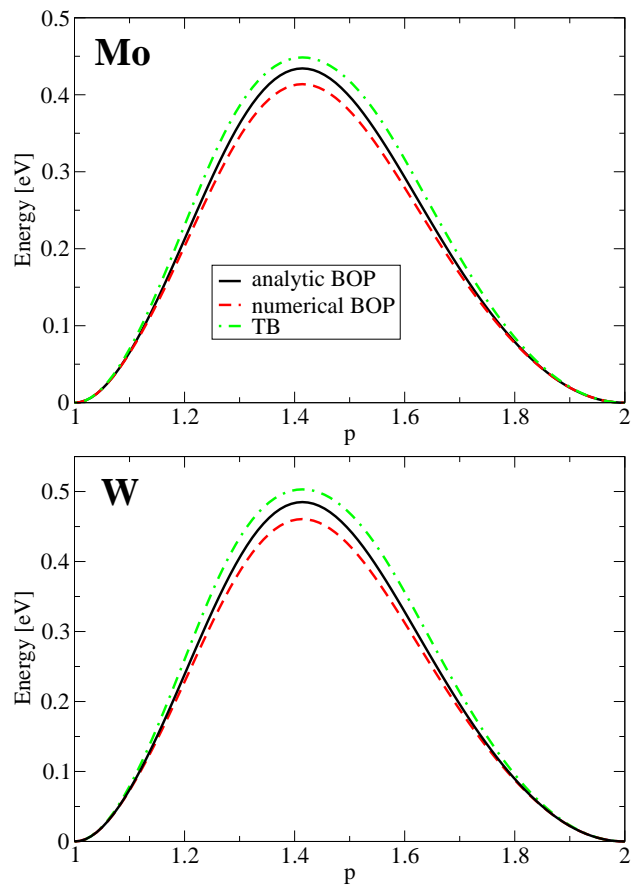

\centering
\begin{tabular}{c}
\epsfig{file=10.eps,width=0.95\linewidth,clip=} \\
\epsfig{file=11.eps ,width=0.95\linewidth,clip=} \\
\end{tabular}
\caption{Binding energy along the orthorhombic deformation path. The minima at $p$ = $1$ and $p$ = $2$ correspond to bcc, 
         the maximum at $p$=$\sqrt2$ to bct.}
\label{fig:DefOrtho}
\end{figure}

The orthorhombic deformation path connects two bcc structures with one symmetry dictated maximum that corresponds to a body-centred 
tetragonal (bct) lattice. This deformation is described by a rotation of the coordinate system to [110], [$\bar{1}$10] and [001], 
respectively. Then the bcc structure is simultaneously elongated along [001] and compressed in the [110] direction. The non-vanishing 
components of the corresponding Lagrangian strain tensor are 
\begin{eqnarray}
\label{eq:Ortho}
 \epsilon_{11}=\frac{p^{-1}-1}{2}, \quad \epsilon_{33}=\frac{p-1}{2} \, .
\end{eqnarray}
Values of $p=1$ and $p=2$ correspond to bcc, $p=\sqrt{2}$ to the bct structure. Our results shown in Fig.~\ref{fig:DefOrtho} show very 
good agreement between analytic and numerical BOP and TB.

\subsection{Point defects}
 
\begin{figure}[htb]
\centering
\begin{tabular}{cc}
\epsfig{file=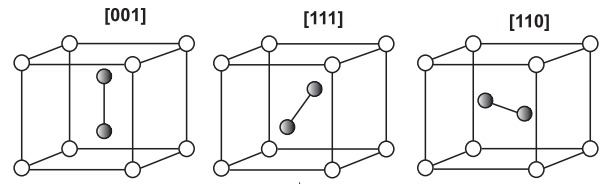,width=0.9\linewidth} &
\end{tabular}
\caption{Self-interstitial atom configurations in bcc.}
\label{fig:SIA}
\end{figure}

We compare the formation energies of (i) a single vacancy in bcc and (ii) self-interstitial atoms (SIAs) in bcc. The SIAs are labelled 
as [001], [111] and [110] according to the Miller indices of the corresponding crystallographic direction as shown in Fig.~\ref{fig:SIA}. 
The sequence of energetic stability of the SIAs in bcc transition metals was identified only in recent years. Ackland and 
Thetford~\cite{Ackland-Thetford} have found (using the semi-empirical Finnis-Sinclair potential) the [110] configuration to be 
most stable for all bcc TMs with the exception of W. Later on, Han {\it et al.}~\cite{Han} predicted on the basis of 
density-functional theory (DFT) calculations
the [111] configuration to have the lowest formation energy for Mo and V. For iron, the [110] SIA is most stable according 
to DFT~\cite{Domaine_Fe, Fu_Fe} and TB calculations~\cite{Duc_Stoner_Fe}. Nguyen-Manh {\it et al.}~\cite{Duc_vac_int} and Derlet 
{\it et al.}~\cite{Duc_potentials} have undertaken a systematic DFT study of SIA for all 5B and 6B group bcc transition metals, with the conclusion 
that in all cases the [111] SIA is the most stable defect. 
This discrepancy between DFT and empirical potentials is related to the binding behaviour at short distances: when the metallic 
material is isotropically compressed, the kinetic energy of the electrons and the ion-ion repulsion increases. In most of the semi-empirical 
schemes this is accounted for only by adjusting the pairwise potential, which is then overestimated and gives rise to a steep 
increase at short interatomic distances. In SIA configurations, however, short bond lengths are present without the corresponding 
significant change in volume. This leads to the discrepancy in the formation energies of interstitials, as pointed out by Han {\it et al.}~\cite{Han}.
The TB model employed here has limitations in describing the short-range interaction appropriately, as pointed 
out earlier~\cite{Mrovec-07-2}. 

\begin{figure}
\centering
\begin{tabular}{cc}
\epsfig{file=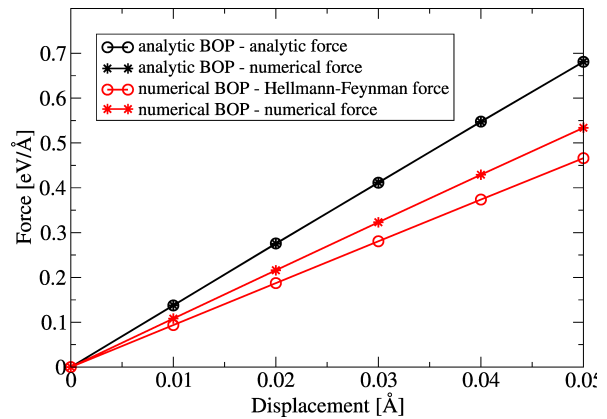,width=0.9\linewidth} &
\end{tabular}
\caption{Absolute values of the {\em x}-component of the force on the central atom in the bcc unit cell as a function of the 
displacement of the atom in {\em x}-direction.}
\label{fig:forces}
\end{figure}

For the SIAs calculations we converged the energies w.r.t. the cell size. For both vacancy and interstitials 
we used a 6$\times$6$\times$6 bcc supercell with 431 atoms for the vacancy and 433 atoms for the SIAs. Our results using analytic BOPs,
numerical BOPS, and TB are compiled and compared with experimental data and with DFT results of Nguyen-Manh {\it et al.}~\cite{Duc_vac_int} 
in Tab.~\ref{tab:SIA}. 
\begin{table}[htb]
\begin{center}
\begin{tabular}{c c c c c c c}
\hline\hline
     &           & expt~\cite{Kittel,Landolt-Boernstein}  &  DFT~\cite{Duc_vac_int} & analytic & numerical & TB\\
     &           &                                        &                         & BOP      & BOP       &   \\
\hline
Mo   &  vac    &  2.6-3.2 &  2.96  & 2.59  & 2.43 & 2.63 \\
     &  [111]  &          &  7.42  & 8.70  & 7.92 & 8.37 \\
     &  [110]  &          &  7.58  & 6.48  & 6.28 & 6.41 \\
     &  [001]  &          &  9.00  & 9.54  & 8.59 & 9.31 \\
     \hline
W    &  vac    &  3.5-4.1 &  3.56  & 4.15  & 3.98  & 4.17\\
     &  [111]  &          &  9.55  & 11.92 & 10.81 & 11.45 \\
     &  [110]  &          &  9.84  & 9.28  & 9.17  & 9.08 \\
     &  [001]  &          &  11.49 & 12.63 & 11.71 & 11.97\\
\hline\hline    
\end{tabular}
\caption{Formation energies of vacancies and interstitials in bcc Mo and W in units of [eV] obtained
from experiment~\cite{Kittel,Landolt-Boernstein}, DFT~\cite{Duc_vac_int} and BOP/TB calculations.}
\label{tab:SIA}
\end{center}
\end{table}
The differences between numerical and analytic BOP for the SIA formation energies are of two origins. First, there are  
differences in the total energy for the same atomic configuration, as illustrated in Sec.~\ref{sec:StructuralStability} 
and~\ref{sec:TransformationPaths}. Second, there are differences in the relaxed structures of the SIA configurations as a consequence
of differences in the forces for the same atomic configuration. 
In order to illustrate the difference between the computed forces we determine the forces using the analytic BOP 
and the numerical BOP formalism and compute the numeric derivative of the energy.
We evaluate the force on a central atom of a two-atom bcc unit cell for different shifts along the x-axis by up to 0.05~\AA\ as
summarised in Fig.~\ref{fig:forces}. 
The numerical forces were obtained using centred finite differences with steps of $\Delta = \pm 10^{-6}$~\AA. 
For the numerical BOPs we observe a significant deviation of the approximate Hellmann-Feynman forces and the numerical forces. 
This inconsistency is the origin of the comparably large deviations of the numerical BOP from the TB results of SIA formation energies and a limitation for the
application of numerical BOPs in dynamic simulations~\cite{Cereceda-2013}. 
For the analytic BOP we find an exact agreement of the analytic and numerical forces. This illustrates 
that the forces in the analytic BOP formalism are strictly consistent with the derivative of the binding energy. 
The consistent treatment of energy and forces in the analytic BOP, together with the linearly-scaling computation of energy and forces, 
enables large-scale molecular-dynamics simulations.

\subsection{Phonons}

\begin{figure}
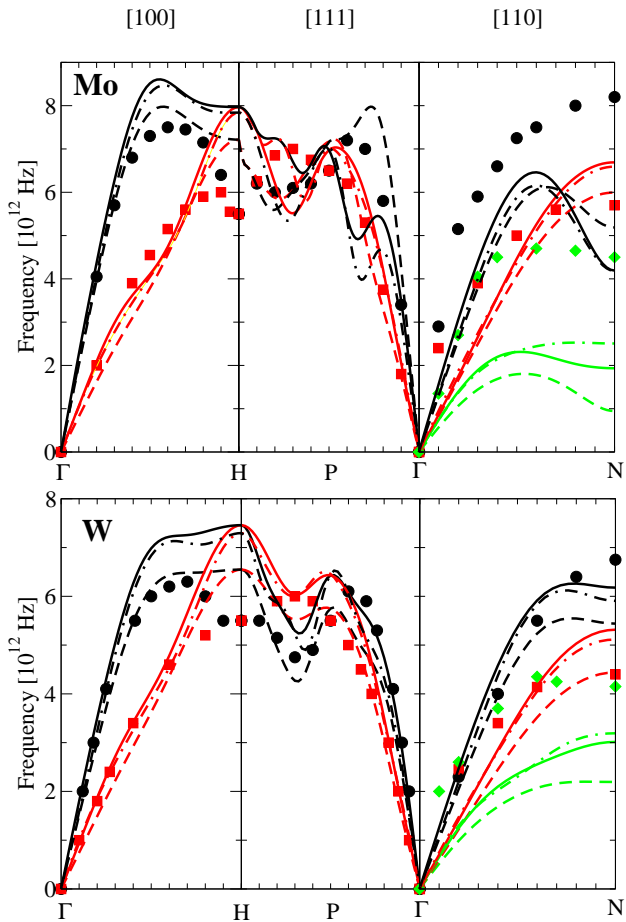

\centering
\begin{tabular}{cc}
\epsfig{file=14.eps,width=0.95\linewidth,clip=} \\
\epsfig{file=15.eps,width=0.95\linewidth,clip=} 
\end{tabular}
\caption{Phonon dispersion curves. Symbols represent experimental data~\cite{Mo-Nb_phonons,W_phonons}, full lines the analytic BOP,
         dashed lines the numerical BOP and dot-and-dash lines TB. Black, red and green represent longitudinal, transversal (T1) and 
         transversal (T2) vibration modes, respectively.}
\label{fig:Phonons}
\end{figure}

We furthermore calculated the phonon dispersion curves for Mo and W and compare our results to the available experimental data. 
We use 216-atom supercells and the Phon software~\cite{Phon} that employs the small displacement method. Our setup 
ensures that the values of the force constant matrices vanish for atoms that are distant from the displaced atom. Our calculated 
phonon dispersion curves for three high-symmetry directions in the Brillouin zone of bcc, $\Gamma$-H, $\Gamma$-N and $\Gamma$-P-H, 
are shown in Fig.~\ref{fig:Phonons}. 
The Cartesian coordinates in reciprocal space of the high-symmetry points are: $\Gamma$=(0, 0, 0), H=(0, 1, 0), N=(0.5, 0.5, 0) 
and P=(0.5, 0.5, 0.5) in units of 2$\pi/a_0$, where $a_0$ is the lattice parameter. We find good overall agreement of the TB and 
BOP calculations with the experimental data. The most considerable deviation is the transversal T2 mode that is too soft in both
Mo and W. This deviation can be reduced by introducing screened bond-integrals to the TB model~\cite{Mrovec-04}.
Comparing the BOP results, we find that the analytic BOP follows the TB results more closely than the numerical BOP. 
The difference between analytic and numerical BOP can be tracked down to the difference in forces on atoms that are used to 
construct the force constant matrices in the small displacement approach that we used to determine the phonon dispersion curves.

\section{Conclusions}

We present a detailed comparison of numerical and analytic bond-order potentials (BOP) based on established BOP parametrisations 
for the bcc refractory metals Mo and W. 
We find that both BOP formalisms capture the electronic density of states in good agreement with TB, in line with previous works. 
We also find good overall agreement of numerical and analytic BOP for the calculation of binding energies, aside from small 
deviations due to the numerical integration scheme in the numerical BOP. 
Despite the good agreement for the bcc ground-state properties, for the sequence of structural stability and for crystallographic 
transformation-paths, we find that the binding energies calculated with analytic BOP tend to agree slightly better with the TB 
results than the numerical BOP. The situation is different in our comparison for point defects and phonon spectra, i.e. for situations where atomic 
forces play an important role. While the forces in the analytic BOP formalism are strictly consistent with the derivative 
of the binding energy, this is not true for the numerical BOPs. For this reason we find that the analytic 
BOPs provide a better agreement with the TB results for point defects and phonon spectra than the numerical BOPs. 

\begin{acknowledgments}
We acknowledge financial support through ThyssenKrupp AG, Bayer MaterialScience AG, Salzgitter Mannesmann Forschung GmbH, 
Robert Bosch GmbH, Benteler Stahl/Rohr GmbH, Bayer Technology Services GmbH and the state of North-Rhine Westphalia as well as
the European Commission in the framework of the ERDF.
\end{acknowledgments}

\bibliography{article}

\end{document}